# Cracking the Quantum Scaling Limit with Machine Learned Electron Densities

Joshua A. Rackers[1*], Lucas Tecot[2], Mario Geiger[3], Tess E. Smidt[4,5]

[1] Center for Computing Research, Sandia National Laboratories
[2] Department of Computer Science, University of California, Los Angeles
[3] Institute of Physics, École Polytechnique Fédérale de Lausanne
[4] Computational Research Division, Lawrence Berkeley National Laboratory
[5] Department of Electrical Engineering and Computer Science, Massachusetts Institute of Technology

*Corresponding author. Email: jracker@sandia.gov

**Abstract:**

A long-standing goal of science is to accurately simulate large molecular systems using quantum mechanics. The poor scaling of current quantum chemistry algorithms on classical computers, however, imposes an effective limit of about a few dozen atoms on current electronic structure calculations. We present a machine learning (ML) method to break through this scaling limit. We show that Euclidean Neural Networks can be trained to predict molecular electron densities from limited data. By learning the electron density, the model can be trained on small systems and make accurate predictions on large ones. This ML model can break through the quantum scaling limit and calculate the electron density for thousands of atoms with quantum accuracy.

# 1  Introduction

One of the grand challenges of science is to simulate large molecular systems like solvated biological macromolecules from first principles quantum physics. The ability to perform such *ab initio* simulations would enable us to predict protein-drug binding, analyze the behavior of materials, and design new enzymes. Unfortunately, large quantum chemistry calculations like these are currently impossible. This is not because the laws of quantum mechanics are not well understood; it is because the algorithms with which they are implemented on classical computers scale sharply with increasing system size. For instance, the so-called "gold standard" method of quantum chemistry, coupled cluster, scales as $O(N^6)$ or $O(N^7)$ with system size. In practical terms this means that while a coupled cluster calculation on a single water molecule might take only a few minutes, an equivalent calculation on a small protein would take several ages of the universe. This behavior defines the scaling limit of about a few dozen atoms for rigorous quantum chemistry.

Despite the scaling of quantum chemistry algorithms, there is no fundamental rule that stipulates that large-scale quantum mechanics calculations cannot be done. The origin of the scaling problem is electron correlation. Conflicting definitions of electron correlation abound, but we adopt a simple definition in this context: the influence exerted on the position of one particular electron by all other electrons in the system. However, we know both empirically and theoretically that this phenomenon of electron correlation cannot have infinite range. The principle of the "nearsightedness of electronic matter" introduced by Walter Kohn in 1996, states that there must be some scale at which electron correlation approaches zero.(*1, 2*) In other words, there exists some (system dependent) distance where quantum mechanics is not needed to compute the force that a given electron exerts on a distant atom.

In this paper we present a machine learning (ML) model that exploits the nearsightedness of electronic matter to perform *ab initio* calculations on systems beyond the scaling limit of quantum chemistry algorithms. The model is trained to predict the electron density, one of the most fundamental quantum properties of a molecular system. Groundbreaking work established that it is possible to machine learn an electron density and subsequent work has highlighted the potential for machine learned density models.(*3–11*) By using a new class of machine learning algorithm, Euclidean Neural Networks, we can train a model on small clusters of molecules and make accurate predictions on large clusters. A recently published, exhaustive database of water clusters makes an excellent test case.(*12*) The performance of our model demonstrates that not only is accuracy on large clusters empirically possible, but there also exists a maximum training cluster size beyond which the accuracy of the model does not increase. This suggests the existence of a "radius of electron correlation" which the ML model can find and exploit to make possible arbitrarily large, systematically improvable electron density predictions. In this way, the model presented here is more than just a surrogate for quantum chemistry. Here, machine learning gives us a unique tool to examine the nature of electron correlation in condensed phase molecular systems. We show that electron densities are uniquely suited for this task of cracking the quantum scaling limit. And we conclude by demonstrating high-level, correlated quantum chemistry electron density predictions on systems of more than 1,000 atoms, well beyond the quantum scaling limit.

## 2 Results

### 2.1 The importance of equivariance

The basic task of our model is to learn the *ab initio* electron density distribution surrounding a set of molecules. An efficient and well-established way to represent this density is to expand it as a linear combination of atom-centered gaussian basis functions,

$$\rho(r) = \sum_{i=0}^{N_{atoms}} \sum_{k=0}^{N_{basis}} \sum_{l=0}^{l_{max}} \sum_{m=-l}^{l} C_{iklm} Y_{l,m} e^{-\alpha_{ikl}(r-r_i)^2} \qquad (1)$$

where $Y_{l,m}$ are the real spherical harmonics, $\alpha_{ikl}$ are the gaussian widths of each basis function, and $C_{iklm}$ are the coefficients for each basis function on each atom.(*13*) Previous work has shown that standard quantum chemistry "density fitting" basis sets which specify a list of $Y_{l,m}$ and $\alpha_{ikl}$ for each element work well for representing an *ab initio* density.(*5*) By choosing a density fitting basis set, our machine learning model is trained simply to predict the $C_{iklm}$ coefficients computed by quantum chemistry. Because each basis function contains a spherical harmonic, each $C_{iklm}$ coefficient is part of a geometric tensor which represents a 3D contribution to the electron density. For instance, for an

$l=1$, or *p*-type, basis function there are three $C_{iklm}$ coefficients that represent the x, y, and z components of that basis function's contribution to the density.

Learning the coefficients of geometric tensors is a unique task for machine learning. Objects in 3D space have certain symmetries; namely they are invariant to permutations and translations, and equivariant to rotations. Equivariance is an intuitive concept: if the input coordinates of a molecule are changed by some arbitrary rotation, the various basis functions that compose its density should rotate with it. This symmetry, however, is not a feature of most current machine learning algorithms. We have recently developed a machine learning framework, called Euclidean Neural Networks (e3nn), which encodes all the symmetries of 3D Euclidean space, including equivariance.(*14–16*) Recent work has highlighted the importance of this property for 3D learning problems. The AlphaFold 2, RoseTTAFold, and ARES models for protein and nucleic acid structure prediction use equivariant machine learning algorithms.(*17–19*) Our e3nn framework has been shown to reduce the amount of training data needed for 3D data by a factor of 1000, compared with models that don't include symmetry.(*20*) And previous electron density learning work using Gaussian Process Regression, another type of machine learning algorithm, has shown equivariance is a desired property for density predictions.(*4*) We set out to test the importance of equivariance for this task of neural network-based electron density prediction.

In figure 1 we show that equivariance is required for an electron density model expressed in an atom-centered basis. In this experiment, we overtrained an equivariant e3nn model on the electron density of a single water molecule. Then we tested that model on rotated versions of the same molecule. We then did the same for an invariant model. Invariant machine learning models are ubiquitous in the machine learning for chemistry field. They guarantee that the outputs of a model do not change with translation or rotation of a molecule. Figure 1 clearly shows that for this electron density learning task, invariance is not sufficient; equivariance is required. The invariant model produces large, rotation-dependent errors on rotated versions of the water molecule. Because the density representation contains vectors and higher rank tensors, the equivariant model produces identical densities for identical objects, while the invariant model does not.

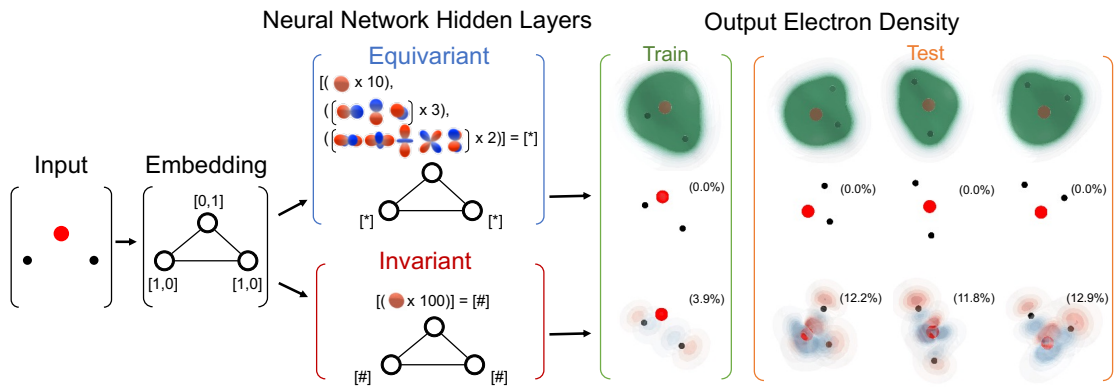

**Figure 1: The importance of equivariance.** Two models, one invariant and one equivariant, are trained on the electron density of a single water molecule. The input one-hot embedding scheme is the same for both models. Both graph neural networks have 3 hidden layers. The invariant network has 100 scalars per node. The equivariant network has 10 scalar, 3 vector, and 2 rank-two tensor features per node. Both networks are tested on three randomly rotated versions of the input water molecule. The errors in the predicted electron density for each are plotted, with red and blue indicating positive and negative errors, respectively. Density difference error, calculated according to equation 2, is listed in parenthesis. The equivariant network achieves precisely zero error on all members of the test set, while the invariant model does not.

Beyond this simple example, we set out to find if equivariance confers any practical advantage in data efficiency for a real problem. To test this, we set up a learning problem on electron densities of clusters of 10 water molecules. The e3nn framework achieves equivariance by imposing that the features at every hidden layer of the neural network be direct sums of irreducible representations. In 3D space, these can be interpreted as spherical harmonics. For practical construction of networks, the user selects a highest angular frequency or degree, $l$, for hidden layer features. Previous work has indicated that the presence of $l>0$ features can increase the data efficiency of neural networks.(20) We set out to test the extent to which this is true for electron densities. We trained five different, equal size, neural networks with $l_{max}$ of 0, 1, 2, 3, 4, and 5. The results, plotted in figure 2, show a dramatic increase in data efficiency from including $l_{max}>1$ features in the network. A network with $l_{max}=2$, for instance, achieves a density difference of 0.44% with just 100 training samples, while a $l_{max}=0$ network requires 100 times more data, 10,000 training samples, to reach similar accuracy. This behavior underscores not only the need for equivariance in the task of learning electron densities, but the advantage that comes from having non-scalar features in the neural network.

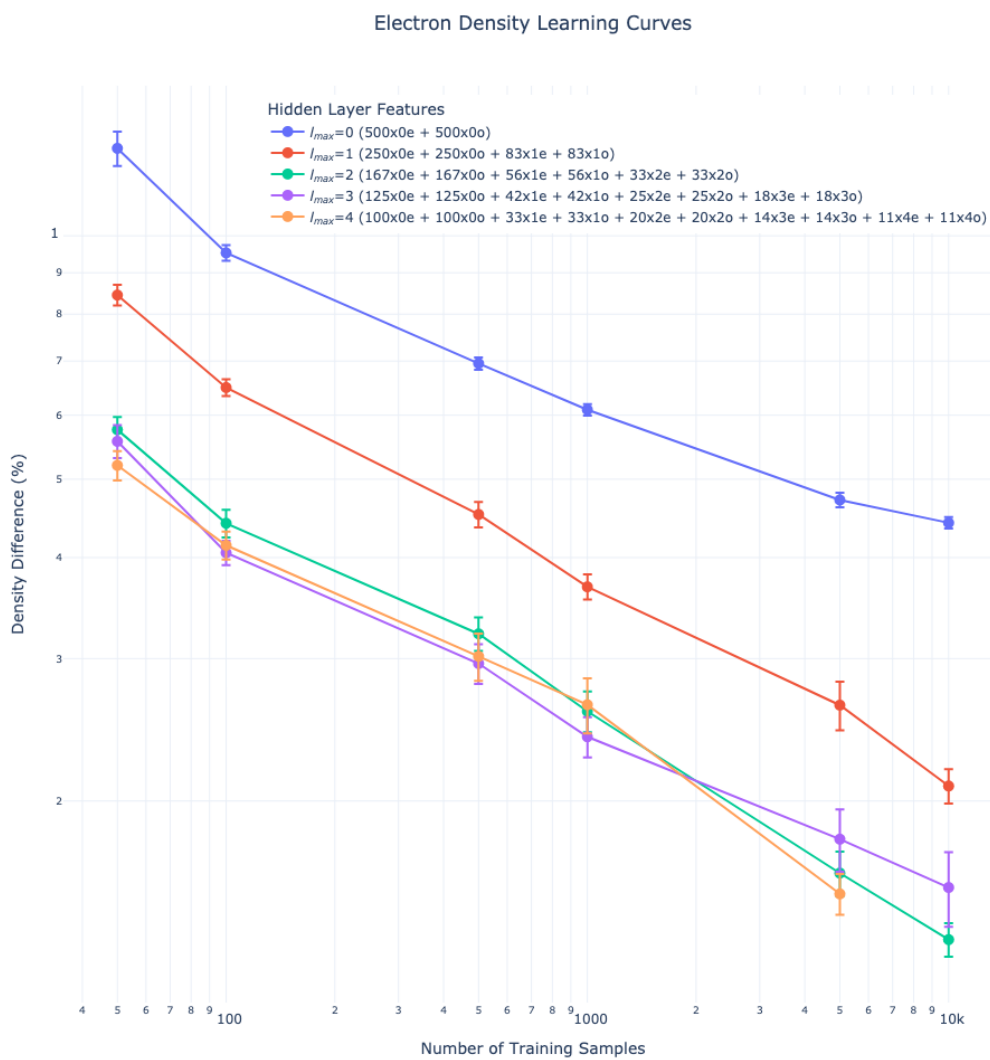

**Figure 2: The importance of equivariant features.** Learning curves on the n=10 water cluster database for e3nn networks with hidden layer features of increasing degree, $l_{max}$, are plotted on a log-log scale. Each network contains equal numbers of features. All curves show characteristic log-log linear dependence for number of training samples vs. density difference error. Both the slope and offset of the learning curves improved by adding $l_{max}>0$ features, up to $l_{max}=2$. (The notation for specifying the hidden features is: "multiplicity x $l$ parity". For example, "33x1e" denotes 33 $l$=1 features with even parity)

## 2.2 Demonstration with Density Functional Theory

One of the fundamental challenges of attempting to perform large-scale quantum chemistry calculations with a machine learning model is the ability to confirm that a prediction is correct. An ML prediction of a CCSD electron density of a 300-atom system is impossible to verify. Therefore, we choose to first do experiments with Density Functional Theory (DFT), a more computationally feasible method. We chose to use the

PBE0 density functional because it produces the most accurate (closest to coupled cluster) electron densities for small test systems.(*21*)

For all experiments presented in this paper, we use the Database of Water Cluster Minima, assembled by Rakshit and co-workers.(*12*) Water clusters make an ideal system of study here because they isolate the problem of learning intermolecular interactions from the problem of learning chemical diversity. The Database of Water Cluster Minima contains structures of all known energy minima of water cluster sizes 3-30. This range of cluster sizes spans the spectrum from gas-phase-like small clusters, up to the bulk-solution-like clusters of n=25-30. All target electron density calculations were performed with the psi4 quantum chemistry package (see Methods and Supplementary Materials).(*22*) The scheme of the following experiments is to train an e3nn model on structures of small clusters, and then test the model on the largest clusters.

### 2.2.1. Experiment 1: The effect of cluster size on density prediction

For our first experiment, we set out to investigate the effect of training set cluster size on the predicted electron densities of the largest clusters available, 30 water molecules. We chose to start at a cluster size of n=7, since lower cluster sizes all have fewer than 100 structures. We trained separate models on cluster sizes n=7, 8, 10, 12, 15, 20, and 25, and then evaluated those models on predictions of the electron density of n=30. The quality of the prediction on each structure was calculated using the standard measure of density difference,

$$\epsilon_p(\%) = 100 \frac{\int dr |\rho_{QM}(r) - \rho_{ML}(r)|}{\int dr \, \rho_{QM}(r)} \quad (2)$$

where and $\rho_{QM}$ is the target density, $\rho_{ML}$ is the predicted density, and the integral is evaluated on a grid of spacing 0.1 Å. A density difference value of 0% indicates identical densities. As a point of reference, superimposing spherical, isolated atom densities typically yields errors of about 20%.(*4*) For each training set, we made sure to include identical numbers of atoms to normalize the total number of samples in the set.

The results in figure 3 show a striking behavior. For low cluster sizes (n=7-10), the performance on n=30 predictably increases with increasing cluster size. However, starting at cluster size n=12, the performance on n=30 converges. This means that the Euclidean Neural Network can accurately learn the electron density of any given atom from an environment of just 12 molecules and that this function is virtually unchanged by the addition of more molecules. Even training with n=25, a cluster size generally assumed to have more bulk solution properties like n=30, does not improve the performance of the model on n=30. In this experiment, the model is unable to find any signal of density contributions to a given atom outside of the average radius of a 12 water molecule cluster, about 6.8 Å.

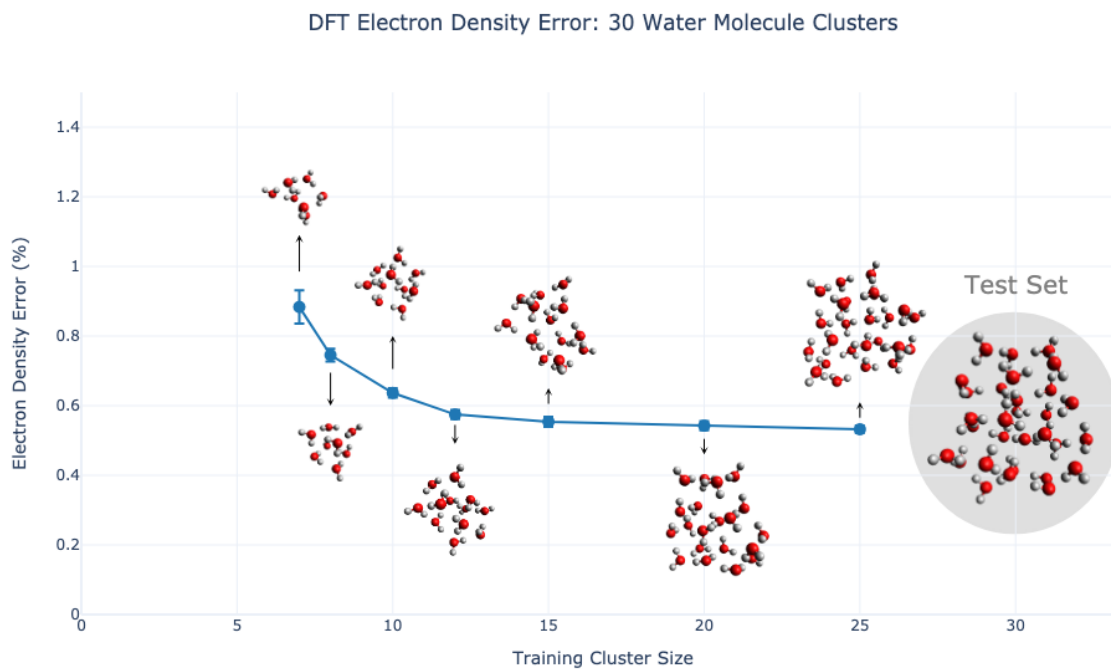

**Figure 3: Convergence of electron density prediction with training cluster size.** Identical e3nn models are trained on electron densities of increasing water cluster sizes. Each model is then tested on predicting the electron densities of the large n=30 water cluster test set. Density difference error on the n=30 set is plotted vs. training cluster size. Example clusters are shown for the training and test sets. The density prediction on n=30 converges around a training cluster size of n=12. (The average cluster radii are 5.5, 5.8, 6.1, 6.8, 7.8, 8.0, 9.7 and 10.6 Å for n=7, n=8, n=10, n=12, n=15, n=20 and n=25, respectively.)

The converged models trained in this experiment achieve average density differences of below 0.6%, on par with the current state-of-the-art in electron density prediction. On the n=12 training set, this accuracy is achieved with fewer than 300 training clusters. As a point of reference, this error is smaller than the error of the PBE density functional vs. Full Configuration Interaction (FCI) on $H_2$ in the complete basis set limit (see Supplementary Materials).(*23*) Accuracy matters because the density can be used to compute properties. For example, the electrostatic potential, calculated from the ML electron density, can be used to quantify hydrogen bonding or drug binding energies. Using the model density trained with cluster size n=12, we computed the electrostatic potential at the van der Waals 0.002 e⁻/bohr³ isodensity surface of every n=30 structure. An example of the accuracy of this electrostatic potential energy surface is shown in figure 4. The average root mean squared error across all structures is 0.006 a.u. This is comparable to state-of-the-art deep neural networks trained directly on the electrostatic potential surface itself.(*24*)

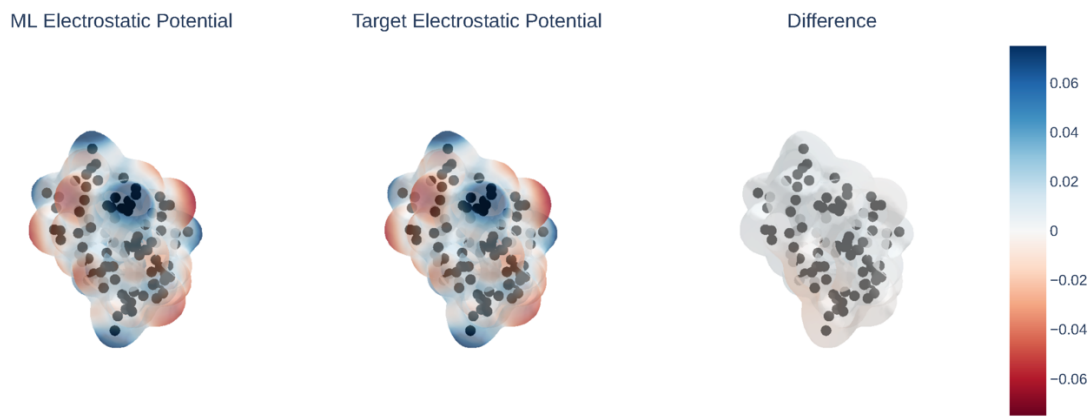

**Figure 4: Electrostatic potential of the e3nn electron density network.** The electrostatic potential of the network trained on n=12 water clusters is compared to the target electrostatic potential on a structure from the n=30 test set. The difference between the target and ML electrostatic potentials is shown on the right. Scale is in atomic units.

### 2.2.2. Experiment 2: The effect of cluster size on force prediction

The majority of the work in machine learning for quantum chemistry to date has focused on energies and forces of molecular structures.(*25*) In our second experiment we set out to see if training an e3nn model on energies and forces would produce the same cluster size convergence behavior as the electron density learning model. The architecture of the model tested is identical to the one used in Experiment 1. The only difference is the outputs of this model are energy and forces as opposed to coefficients of basis functions. ML forces are calculated as atomic position derivatives of the energy using neural network backpropagation, ensuring that the learned forces are conservative.(*26*) We evaluated this model by comparing the predicted force vs. the reference force for the n=30 water cluster test set.

The results, plotted in figure 5, show that training on energies and forces does not produce the same behavior as training on densities. The accuracy of the resulting models is excellent. The model trained on the n=25 clusters, achieves a mean absolute error of 0.1 kcal/mol/Å. This error is comparable to state-of-the-art ML force fields, with an error of less than 1% on the 10.2 kcal/mol/Å average total force magnitude of the n=30 test set.(*20*) However, in this case there is no convergence as is seen with densities. The mean absolute error in forces on the large system decreases progressively with system size. This, of course, does not mean that convergence will never occur. It does mean, however, that it has not been reached by n=30 for this system.

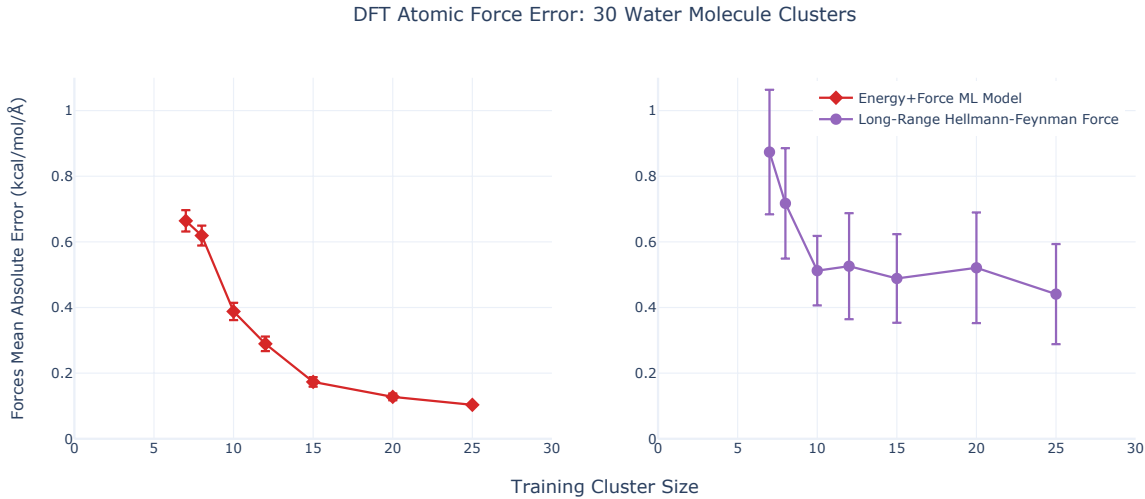

**Figure 5: Atomic force prediction with increasing training cluster size**. (Left) Identical e3nn models are trained on energies and forces of increasing training cluster sizes. Each model is then tested on predicting the atomic forces of the n=30 water cluster test set. Mean absolute error (MAE) in the forces is plotted vs. training cluster size. No convergence behavior is seen. Increasing the size of the training cluster consistently improves the prediction of n=30 forces. (Right) Long-Range Hellmann-Feynman forces for each training cluster size are calculated from the corresponding ML electron density model. The Hellmann-Feynman forces exhibit the same convergence behavior as the densities themselves.

A compelling reason to learn the electron density is that one of the many properties that can be derived from it are forces, via the Hellmann-Feynman Theorem. The Hellmann-Feynman Theorem states that given an electron density, atomic forces can be calculated exactly from simple electrostatics.(*27*) In practice, however, the Hellmann-Feynman Theorem is not valid for short-range forces with basis sets like the aug-cc-pVTZ basis set used in this work.(*28–31*) It is perfectly valid, however, to compute long-range forces with the Hellmann-Feynman Theorem. To test the behavior of these long-range forces with increasing training set cluster size, we computed the long-range contribution to the Hellmann-Feynman forces (see Methods and Supplementary Materials for details) from the predicted densities of the models trained in Experiment 1. These results are also plotted in figure 5. Like the ML forces, the long-range Hellmann-Feynman forces calculated from the ML electron densities of Experiment 1 are very accurate. For example, the long-range Hellmann-Feynman forces calculated from the ML electron density model trained on the n=12 dataset give a mean absolute error of 0.5 kcal/mol/Å on the n=30 test set. This is an error of 3% on the average long-range force magnitude of 16.0 kcal/mol/Å. Note that the average total magnitude of the forces in the long-range case is larger than for the total forces because the structures in the database are close to DFT minima.

Most importantly, figure 5 illustrates that the long-range forces computed from the electron density model follow the same behavior with respect to cluster size as the densities themselves. The accuracy of the prediction on 30 water molecule clusters converges around a training cluster size of n=12. The convergence of these long-range forces corroborates the main observation: equivariant neural networks trained on the electron density converge much more rapidly with cluster size than models trained on energies and forces.

To quantify the convergence behavior in Experiments 1 and 2 we set out to estimate the derivative of the error vs. training cluster size plots in figures 3 and 5. We fit the data with a simple decaying exponential function (see Supplementary Materials for fits) and plot the derivative of each function in figure 6. The derivatives with respect to average training cluster radius show a clear trend. The ML Electron Density and Long-Range Hellmann-Feynman Forces display similar convergence behavior with distance and both converge faster than the ML Energy+Force model.

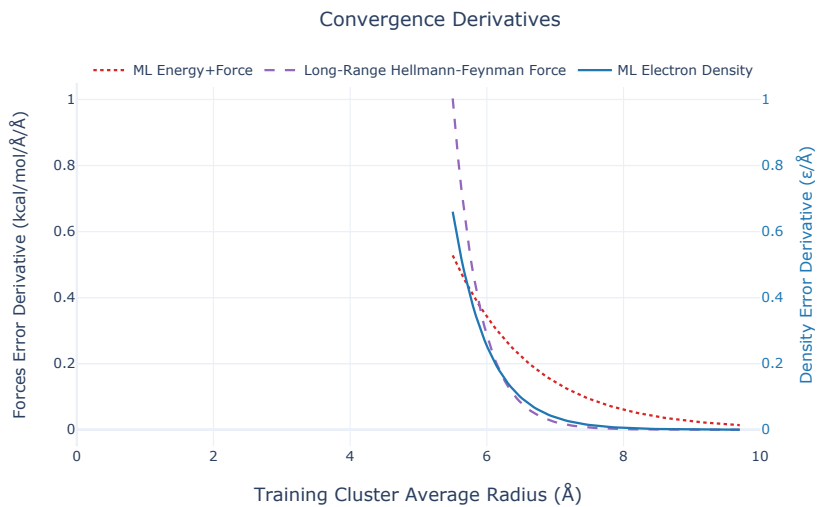

**Figure 6: Derivatives of error vs. training set cluster average radius.** The ML electron density error (right axis), ML force error (left axis), and long-range Hellmann-Feynman force error (calculated from the ML electron density, left axis) are all fit with a decaying exponential function. The derivative of each function is plotted vs. the average training cluster radius. The derivatives of the ML electron density and long-range Hellmann-Feynman force plots approach zero at shorter distance than the pure ML force. Alignment of y-axes units is coincidence.

## 2.3 Cracking the Scaling Limit with Coupled Cluster

Coupled cluster methods are the so-called gold-standard of quantum chemistry. They are accurate and expensive to compute because they systematically include the effects of electron correlation across an entire molecular system. The two most commonly used versions, CCSD (coupled cluster with single and double excitations) and CCSD(T) (coupled cluster with single and double perturbative triple excitations), are used as benchmarks against which other approximate methods are measured.(*32*) In many cases coupled cluster is accurate enough to be used as a surrogate for experiment when experimental data is unavailable.(*33*) Because of the $O(N^6)$ or worse scaling, however, it is virtually impossible to perform these calculations on anything more than a few dozen atoms. We endeavored to see if a machine learning electron density model can be used to perform correlated coupled cluster calculations that break through this scaling limit. The evidence from Experiments 1 and 2 suggests that it is possible to train a model on electron densities of small molecular clusters and achieve accurate results on arbitrarily larger systems. Figure 3 indicates that if we can obtain training data for an ML model at a cluster size of about 12 water molecules, we can likely predict the densities of much larger clusters equally well. We set out to investigate if this same behavior holds for CCSD calculations.

### 2.3.1. Experiment 3: Coupled cluster density prediction

In Experiment 3, (just as in Experiment 1) we trained "experiment" ML models on CCSD electron densities of training set cluster sizes n=7-15. Because CCSD calculations on the n=30 cluster size are impossible, we were forced to employ two surrogates to evaluate how well these "experiment" models perform on larger structures.

We first created a separate "reference" set of n=15 CCSD electron density calculations that does not include the data used to train the n=15 "experiment" model. The left side of figure 7 shows the accuracy of each trained "experiment" model on this n=15 CCSD "reference" test set. These n=15 CCSD calculations, however, are lacking as a surrogate for bulk water structure. Therefore, we set out to create a second method to evaluate the quality of the "experiment" models on the n=30 water cluster structures. To do this, we trained a separate ML model on the "reference" n=15 CCSD electron densities. We then compared the predicted electron densities of the "experiment" models against the electron densities of the "reference" model on the structures of the n=30 water cluster database. This allows us to make comparisons on n=30 structures using our best-possible ML(CCSD) estimate. The right side of figure 7 shows this comparison. The curve exhibits the same convergence behavior seen in Experiment 1.

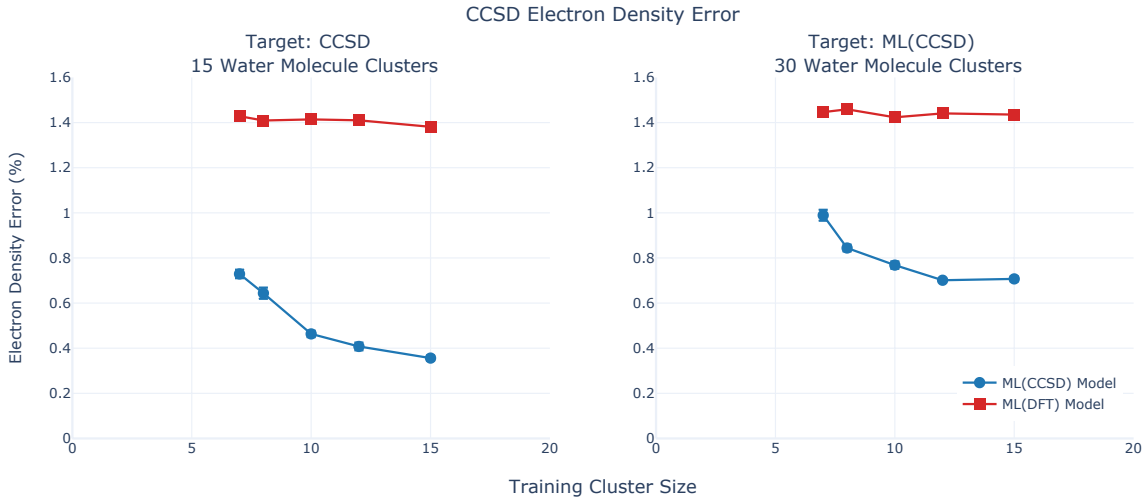

**Figure 7: Convergence of ML(CCSD) electron densities with training set cluster size.** Identical e3nn models are trained on CCSD electron densities of increasing water cluster sizes. Density difference error vs. training cluster size is calculated against two different target electron densities. (Left) Models are evaluated against CCSD electron densities for the n=15 "reference" water cluster database. (Right) Models are evaluated on the n=30 water cluster structures against reference ML(CCSD) electron densities trained on the n=15 "reference" set. Also plotted for both panels is the performance of the trained ML(DFT) models from Experiment 1 against the respective references.

We also tested the ML(DFT) models trained in Experiment 1. We calculated the density difference between the densities predicted by the ML(DFT) and both reference densities. The results show that the machine learning model can clearly distinguish between CCSD and DFT levels of theory. The error of the ML(DFT) with respect to the reference ML(CCSD) model is consistent with the average density difference error of DFT vs. CCSD from quantum chemistry on the n=7 set of about 2%.

We used the ML(CCSD) electron density network from Experiment 3, trained on clusters of n=12 water molecules, to predict the electron densities of structures well beyond the scaling limits of quantum chemistry. In figure 8 we show predictions for clusters of 64 and 6,400 molecules. While a calculation on 64 water molecules is outside of the scaling limit for coupled cluster, it is still possible for DFT. Figure 8 shows that the ML model trained on n=12 water clusters is still accurate for the n=64 structure. Moreover, the ML models are clearly able to distinguish between DFT and CCSD levels of theory, even for this large structure. On the other hand, 6,400 water molecules is beyond the reach of almost any quantum chemistry method. Figure 8 shows, however, that large calculations like this pose no problem for the neural network. While a full CCSD calculation on 6,400 water molecules would take more than the age of the universe, the electron density of this structure can be calculated with near-CCSD accuracy in under one second with our Euclidean Neural Network model on a single graphics processing unit (GPU).

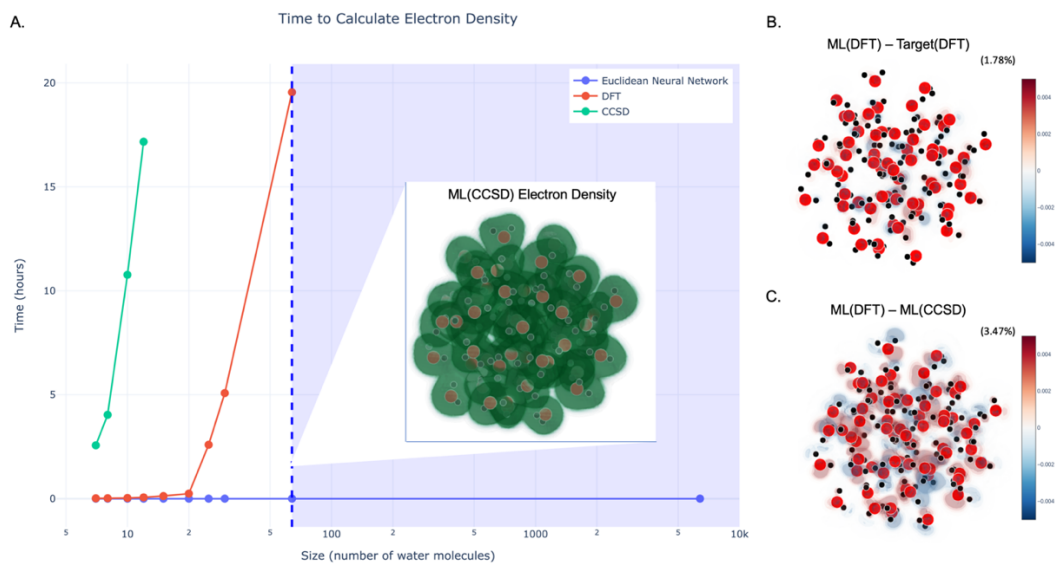

**Figure 8. Machine learned electron densities past the quantum scaling limit.** Panel A shows the time to calculate the electron density of water clusters of increasing size for DFT, CCSD, and our Euclidean Neural Network. The shaded area to the right of the dashed vertical line roughly denotes the area inaccessible to quantum chemistry, past the scaling limit. CCSD calculations in this regime would take more than 100 CPU-years. The Euclidean Neural Network exhibits linear scaling. The insert shows the machine learned CCSD prediction for a 64 water molecule cluster. Panel B shows the density difference map between the ML(DFT) and DFT densities. Panel C shows the density difference map between the ML(DFT) and ML(CCSD) predictions. Values are in $e^-/bohr^3$. In both B and C, the density difference is shown in parentheses. Note that the density difference between the ML(DFT) and ML(CCSD) models is larger than the intrinsic error of ML(DFT) itself, indicating that the ML model is picking up *bona fide* differences between CCSD and DFT.

# 3 Discussion

The experiments in this work probe the ability of Euclidean Neural Networks to learn molecular electron densities. Several interesting observations arise from the results of these experiments.

Equivariance is essential for an accurate machine learning model of electron densities. The data we are predicting in this task are geometric tensors, and the results in figure 1 show that equivariance is a required attribute for sensible results for this data type. Moreover, figure 2 shows that the presence of higher-order ($l>1$) tensors in the hidden layers of an equivariant network confers a large increase in data-efficiency. This suggests that having features in the network that have the same data type as the target output is a natural fit for the learning task.

The practical implication from this work is a blueprint for how to solve the "training data problem" in the machine learning for chemistry field. The goal of most ML for chemistry models is to perform calculations on systems that are too large for *ab initio* calculations. The problem, however, is that it is impossible to generate training data for these large systems. This poses an existential conundrum for condensed-phase molecular systems where long-range forces matter. Omitting long-range forces can produce large errors in molecular dynamics simulations.(*34*) If the training data cannot capture long-range forces, then the model will not be accurate. This work suggests that learning the electron density is a way out of this trap. The data in figures 3 and 7 show that it is possible to train on small clusters of molecules and expect similarly accurate predictions on larger clusters. Importantly, the minimum training cluster radius of 7 Å for water suggests that this strategy is practically viable with current computer hardware.

By restricting the training data used for a given ML model, we were able to articulate a rigorous scientific question: what is the effective radius needed to make accurate predictions of the electronic structure of bulk water? Experiments 1 and 3 directly address this question and show striking convergence behavior. We suggest that this is the result of the neural network picking up an effective "radius of electron correlation" for water clusters. Figures 3 and 7 show that, up to the sensitivity of the model, electrons outside of a radius of about 7 Å, have no measurable effect on the electron density surrounding an atom. It is possible, of course, that this is an artifact of the model. However, the results of Experiment 2 suggest that this is not the case. In Experiment 2, we used an identical neural network architecture, but predicted atomic forces rather than electron densities. In this case, the network does pick up signal from atoms outside of the 7 Å radius observed for electron densities. This strongly suggests that the density-learning network is revealing a real physical feature of the physics of electron correlation.

It may seem curious that the density-learning and force-learning experiments produce differing behavior, but perturbation theory offers a straightforward interpretation of this observation. A toy example illustrates this clearly. An atom, A, 10 Å away from another atom, B, can exert a force on atom B without substantively changing atom B's electronic structure. To compute the force atom A exerts on atom B, the first order contribution is from the electric field generated by A at B. Then, the second and higher order contributions come from the relaxation of the electron density of B in response to the field, field gradient, etc. from A. For well-behaved systems at long-range, the first order effect should dominate, with second and higher order contributions producing diminishing effects. This is exactly in line with rigorous Symmetry Adapted Perturbation Theory.(*35*) Moreover, recent work comparing the many-body expansion convergence for energy versus electron density showed that electron densities converge at lower body-order than energies for molecular clusters.(*36*) Our observation that learned densities converge faster than forces with cluster size is consistent with this result, suggesting that, at least for water clusters, the effects of electron correlation at long range are well-behaved and systematically diminishing.

This study demonstrates that machine learning models can be more than just surrogates. In this case, the ML model doesn't just reproduce quantum chemistry calculations, it tells us something about quantum chemistry itself by revealing a "radius of correlation". Despite the conceptual appeal of the principle of "nearsightedness of electron matter", science has struggled to devise experiments that probe what exactly "nearsighted" means in a specific, unbiased way. The ML models here give us an unbiased tool to explore the nature of electron correlation from data. In a general sense, the model probes something unknown (physical electron correlations) by examining something that is known (statistical correlations in neural networks).

Our results strongly suggest that the same behavior observed for DFT holds for coupled cluster results as well. This is noteworthy because while the density functional used for the ML(DFT) models, PBE0, contains some nonlocal electron correlation, it is not a fully correlated method.(*37*) CCSD on the other hand, contains full non-local electron correlation (up to double excitations). This makes it a more stringent test of the "effective radius of electron correlation" argument. The fact that we see the same behavior in figure 7 as in figure 3, implies that convergence with training set radius was not an artifact of the DFT method. It suggests that even at higher, inaccessible levels of theory like CCSD(T) or full configuration interaction (Full-CI), this trend will hold.

There are limitations to the ML electron density framework presented here. The results are specific to water clusters. The "radius of correlation" suggested here is likely to be highly system dependent. Further work will be required to characterize this limit for more complex systems like biomolecular fragments. Additionally, while this work strongly suggests that equivariant neural networks are a good tool to probe the "radius of correlation", more work will be required to conclusively characterize the spatial effects of electron correlation for more complex systems.

## 4 Conclusions

The electron density is one of the most important physical observables of a molecular system. Unfortunately, computing the electron density for large systems, like biomolecules or condensed phase liquids, with traditional quantum chemistry methods is practically impossible. The work presented here shows that a machine learned electron density model that uses Euclidean Neural Networks can break through this quantum scaling limit. In the case of water clusters, our experiments suggest that this is possible because the network discovers an effective "radius of electron correlation". This observation establishes a framework for machine learning models trained on small systems that can systematically approach previously impossible gold standard quantum chemistry results for arbitrarily large molecular systems. The potential impacts are far-reaching. This framework could be used, for example, to build *ab initio* models of protein dynamics, from a simple training set of protein fragment interactions. These quantum-accurate simulations would revolutionize applications from computational drug design to protein engineering.

# 5 Materials and Methods

## 5.1 Data Generation

All the training and test data used in this study were obtained from quantum chemistry calculations using the psi4 quantum chemistry software package. The structures for these calculations were obtained from the Database of Water Cluster Minima from Rakshit and co-workers.(14)

We performed quantum chemistry calculations at the PBE0 and CCSD levels of theory for this study. For both methods, we calculated energy, atomic forces, and electron density. The electron density is projected into an auxiliary basis set using density fitting (see Supplementary Materials). All PBE0 calculations use the aug-cc-pVTZ basis set and all CCSD calculations use the aug-cc-pVDZ basis set. All input structures and results are publicly available in the online database: https://zenodo.org/record/5563139.

## 5.2 Euclidean Neural Networks

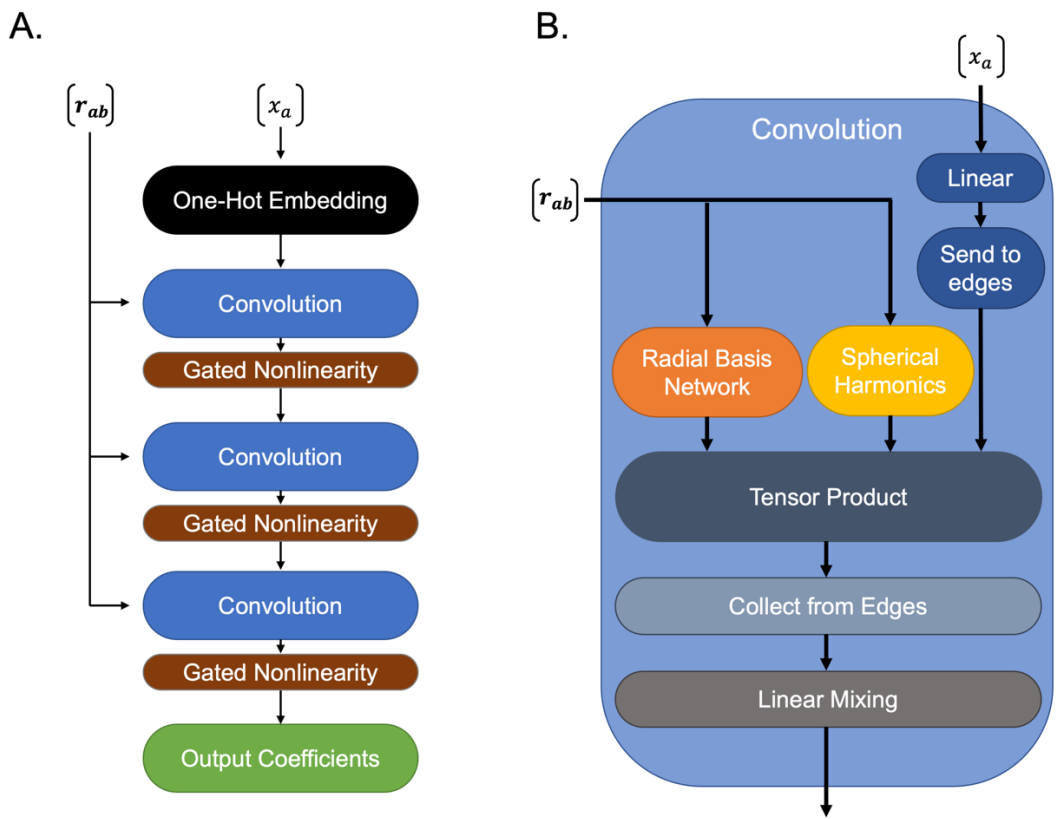

**Figure 9: Architecture of the e3nn electron density networks.** The input features, *x*, and geometry, *r*, are inputs to the network. The output is the coefficients of the electron density in the density-fitting basis.

The network we train is implemented in e3nn.nn.models.gate_points_2101, in version 0.3.5 of the e3nn package.(*38*) A graphical description of the model is shown in figure 9. We set the network to have three hidden layers, each with features defined by irreps = 125x0o + 125x0e + 40x1o + 40x1e + 25x2o + 25x2e + 15x3o + 15x3e. The notation for specifying the hidden features is: "multiplicity x *l* parity". For example, "40x1e" denotes 40 *l*=1 features with even parity. For the convolutions we take spherical harmonics up to $l_{max}$=3. The radial function has a maximum radius of 3.5 Å, approximately equal to the distance of the first hydration shell of liquid water at room temperature. The fully connected neural networks used for the radial basis networks each have one hidden layer with 128 neurons. See Supplementary Materials for model implementation details.

## 5.3 Training Details

For the PBE0 experiments, we used the following training set sizes: 468 n=7, 409 n=8, 328 n=10, 273 n=12, 218 n=15, 164 n=20, and 131 n=25, so that all training sets have ~9,840 atomic environments. The test set is 1000 PBE0 calculations for n=30. For the CCSD experiments, we used the following training set sizes: 468 n=7, 409 n=8, 328 n=10, 273 n=12, and 218 n=15, so that again the number of atomic environments in each training set is equal. There were two test sets used for the CCSD experiments. Both use a separate "reference" set of 700 n=15 CCSD density calculations. The first test set is the "reference" n=15 densities themselves. The second is the electron density prediction of the "reference" model trained on the 700 n=15 CCSD calculations evaluated on the 1000 n=30 structures.

All models were trained for 500 epochs with a learning rate of 0.01 using the Adam optimizer. The metrics reported in the paper are taken from the last 100 epochs of each run, when the performance on the test set has converged. For the electron density learning problems, the loss function is the mean squared error between the learned and target coefficients,

$$Loss_{density} = \frac{1}{n} \sum_{i=1}^{n_{aux}} \left( C_i^{ML} - C_i^{target} \right)^2 \tag{3}$$

where n is the number of basis functions in the density-fitting basis set.

To better normalize the problem, we performed two transformations on the target coefficients before training. First, we normalized each coefficient to be a population. This made the coefficient of each basis function equal to the number of electrons contained in that function. Second, we subtracted away the density of isolated atoms. We performed quantum chemistry calculations on isolated oxygen and hydrogen atoms and then subtracted away this "unperturbed" electron density from the target such that the total number of electrons in the learned density sums to zero. Note that this only affects the $l$=0 functions.

For Experiment 2, the input and structure of the hidden layers are identical to the electron density learning network of Experiments 1 and 3. The only difference is that rather than outputting density coefficients, the network outputs a single scalar value: the energy of the configuration. Previous work has shown that the most accurate and data efficient way to learn a potential energy surface is to include forces in the loss function.(*26*) To ensure conservative forces, atomic forces are calculated by computing the gradient of the network energy with respect to the input atomic positions. To normalize, we subtract $N_{mol} \times E_{iso}^{mol}$ from the target energy. $E_{iso}^{mol}$ is the energy of an isolated water molecule in its equilibrium geometry and $N_{mol}$ is the number of molecules. The total loss function is then defined as:

$$Loss_{PES} = \left(E_i^{ML} - E_i^{target}\right)^2 - \frac{1}{n}\sum_{i=1}^{n_{atoms}}\sum_{j=1}^{3}\left(F_{ij}^{ML} - F_{ij}^{target}\right)^2 \tag{4}$$

where $F$ is the force component on each atom.

## 5.4 Hellmann-Feynman Forces

Long-range Hellmann-Feynman forces were computed according to the equation

$$F_i = Z_i\left(\sum_{i \neq j}\frac{Z_j(\mathbf{R}_j - \mathbf{R}_i)}{|\mathbf{R}_j - \mathbf{R}_i|^3} - \int d\mathbf{r}\,\rho(\mathbf{r})\frac{\mathbf{r} - \mathbf{R}_i}{|\mathbf{r} - \mathbf{R}_i|^3}\right) \tag{5}$$

where $Z$ are the nuclear charges and $\rho$ is the electron density. To compute the long-range contribution, we excluded all contributions to the force from nuclei and basis function centered on nuclei less than 7 Å away. This is necessary to compute converged Hellmann-Feynman forces (see Supplementary Materials).

# Acknowledgements

J.A.R. thanks Shivesh Pathak, Aidan Thompson, Andrew Simmonett, Simon Batzner, and Alex Lee for helpful discussions. L.T. thanks Cho-Jui Hsieh for mentorship support.
**Funding:** J.A.R. and L.T. were supported by the Harry S. Truman Fellowship and the Laboratory Directed Research and Development Program of Sandia National Laboratories. T.E.S. was supported by the Luis Alvarez Fellowship and the Laboratory Directed Research and Development Program of Lawrence Berkeley National Laboratory. Sandia National Laboratories is a multimission laboratory managed and operated by National Technology & Engineering Solutions of Sandia, LLC, a wholly owned subsidiary of Honeywell International Inc., for the U.S. Department of Energy's National Nuclear Security Administration under contract DE-NA0003525. This paper describes objective technical results and analysis. Any subjective views or opinions that might be expressed in the paper do not necessarily represent the views of the U.S. Department of Energy or the United States Government.

**Authors contributions:**
Joshua Rackers: Conceptualization, Methodology, Investigation, Software, Writing – Original Draft
Lucas Tecot: Investigation, Formal Analysis, Resources, Writing – Review and Editing
Mario Geiger: Software, Writing – Review and Editing
Tess Smidt: Methodology, Software, Writing – Review and Editing
**Competing interests:** The authors declare no competing interests.
**Data and materials availability:** All quantum chemistry data is publicly available in an online repository at https://zenodo.org/record/5563139. All code needed to reproduce the experiments and analysis is publicly available on GitHub at https://github.com/JoshRackers/equivariant_electron_density.

# List of Supplementary Materials

Methods

Supplementary Text

# Supplementary Materials for

## Cracking the Quantum Scaling Limit with Machine Learned Electron Densities

Joshua A. Rackers[*], Lucas Tecot, Mario Geiger, Tess E. Smidt

*Corresponding author. Email: <jracker@sandia.gov>

**This PDF file includes:**

> Method Details
> Supplementary Text
> Figures S1-S5

# 1 Method Details

## 1.1 Electron density fitting

The output electron density of the quantum chemistry calculation is the one-particle, reduced density matrix, referred to hereafter as the density matrix. This matrix has size NxN where N is the number of basis functions in the orbital basis used for the quantum chemistry calculation. Because the density matrix scales as the square of the system size, it does not make an ideal target for machine learning predictions on large molecular systems. To fix this, we choose to project the density matrix onto an auxiliary atom centered basis. The size of the density represented in the auxiliary basis scales linearly with system size.

The process of projecting the density matrix on to an auxiliary basis is known as "density fitting", and it is widely used in quantum chemistry. There are specific basis sets that have been developed to minimize the error in this projection. We choose to use one of the most common density-fitting basis sets, def2-universal-JFIT.(*39*) The procedure for performing the projection is as follows.

We wish to represent the total density, $\rho$, in terms of a sum of coefficients, $C$, multiplied by our pre-defined density-fitting basis functions, $\phi$

$$\rho(\boldsymbol{r}) = \sum_i^{n_{aux}} C_i \phi_i^{aux}(\boldsymbol{r}) \tag{1}$$

where $\phi$ are combinations of a spherical harmonic and gaussian, $Y_{l,m} e^{-\alpha_i (r-r_i)^2}$.
To obtain these coefficients, we start from the density represented as a density matrix,

$$\rho(\boldsymbol{r}) = \rho_{ab}(\boldsymbol{r}) = \sum_{ab}^{n_{AO}} D_{ab} \phi_a(\boldsymbol{r}) \phi_b(\boldsymbol{r}) \tag{2}$$

where $\phi_a$ and $\phi_b$ are the atomic orbitals in the atomic orbital (AO) basis, and $D_{ab}$ is the density matrix. We wish to represent the overlap integrals $\phi_a \phi_b$ as an expansion in the auxiliary basis,

$$\phi_a(\boldsymbol{r}) \phi_b(\boldsymbol{r}) = \sum_i^{n_{aux}} d_i^{ab} \phi_i^{aux}(\boldsymbol{r}) \tag{3}$$

where the expansion coefficients, $d$, are composed of the three center overlap integrals of the AO basis with the auxiliary basis,

$$J_{abj} = \int d\boldsymbol{r_1} \int d\boldsymbol{r_2} \frac{\phi_a(\boldsymbol{r_1}) \phi_b(\boldsymbol{r_2}) \phi_j^{aux}(\boldsymbol{r_2})}{r_{12}} \tag{4}$$

and the so-called metric,

$$M_{ij} = \int d\boldsymbol{r_1} \int d\boldsymbol{r_2} \frac{\phi_i^{aux}(\boldsymbol{r_1}) \phi_j^{aux}(\boldsymbol{r_2})}{r_{12}} \tag{5}$$

such that,

$$d_i^{ab} = \sum_j^{n_{aux}} J_{abQ} (M_{ij})^{-1} \tag{6}$$

We can then apply these expansion coefficients to equation 3 to obtain,

$$\rho(\boldsymbol{r}) = \sum_{ab}^{n_{AO}} D_{ab} \sum_i^{n_{aux}} d_i^{ab} \phi_i^{aux}(\boldsymbol{r}) \tag{7}$$

The order of summation can be switched to give,

$$\rho(\boldsymbol{r}) = \sum_i^{n_{aux}} \sum_{ab}^{n_{AO}} D_{ab} d_i^{ab} \phi_i^{aux}(\boldsymbol{r}) \tag{8}$$

Thus, the coefficients we seek, $C$, are obtained from the density matrix by,

$$C_i = \sum_{ab}^{n_{AO}} D_{ab} d_i^{ab} \tag{9}$$

We have implemented this density-fitting procedure using the psi4numpy library.(*40*) The code can be found in the equivariant_electron_density Github repository.

### 1.1.1. Constraining the number of electrons

The density-fitting procedure is not guaranteed to give an integer number of electrons. Therefore, we implemented a Lagrange multiplier scheme laid out in references (*41, 42*) to constrain the total charge in the auxiliary basis. Following equations 15-17 in reference (*42*), we wish to find a Lagrange multiplier, $\lambda$, such that

$$0 = \boldsymbol{J}_{abj} - \boldsymbol{M}_{ij} \cdot \boldsymbol{C}' + \lambda \boldsymbol{q} \tag{10}$$

where $q_i$ is the amount of charge associated with each auxiliary basis function,

$$q_i = \int d\boldsymbol{r}\, C_i \phi_i^{aux}(\boldsymbol{r}) \tag{11}$$

(only basis functions with spherical harmonic $l{=}0$ contribute). The multiplier, $\lambda$, then is given by the equation:

$$\lambda = \frac{Q - \boldsymbol{q} \cdot \boldsymbol{D}_{ab}}{\boldsymbol{q} \cdot (\boldsymbol{M}_{ij})^{-1} \cdot \boldsymbol{q}} \tag{12}$$

The new coefficients are then given by:

$$\boldsymbol{C}' = \boldsymbol{D}_{ab} + (\boldsymbol{M}_{ij})^{-1} \cdot \lambda \boldsymbol{q} \tag{13}$$

These new coefficients define the final output of the density-fitting calculation.

### 1.1.2. Final output: electron density in an atom-centered basis

We list the final, charge-constrained density-fitted coefficients, along with the corresponding basis set information in an output file. These $C'$ coefficients are what is learned by our density-learning networks. They define the $C_{iklm}$ coefficients in equation 1 of the main text.

## 1.2 Euclidean Neural Networks

The data type we want to learn in this study is a geometric tensor of *C_iklm* coefficients. For the basis set used, the tensors go up as high as rank-4 (*f*-type functions). Because we use this particular data type, we need a neural network which respects the symmetries of this data. Namely, we require a network which is invariant to translations and permutations and equivariant to rotations. While neural networks with the first two symmetries are common, equivariant networks have only recently been developed. Here, we use Euclidean Neural Networks, implemented in the e3nn software package, to achieve equivariant electron density prediction.

### 1.2.1. Structure of a Euclidean Neural Network

At a broad level, our e3nn network is structured as a graph convolutional neural network. It has nodes and edges, with each atom defining a node. Input information is given to each node and then that information is passed through hidden layers before outputting coefficients of the electron density. What makes this Euclidean Neural Network unique is that in the hidden layers, the features on nodes transform as irreducible representations of 3D rotations and spatial inversion (the group O(3)). This stands in contrast to the standard "invariant" ML models for chemistry, where features on nodes do not transform under rotation or inversion.

**Input Layer:**

The input is a simple one-hot encoding. Each node is assigned an input feature, $x_a^{(1)}$, [1,0] for hydrogen or [0,1] for oxygen. In the language of e3nn, this corresponds to each node having an input irreducible representation, "irrep", of 2x0e (2 channels with *l*=0 and even parity). This is illustrated in figure 1 of the main text.

**Hidden Layers:**

Each hidden layer of the network is a gated convolution which operates on the features of the preceding layer, $x^{(i)}$,

$$x_a^{(i+1)} = \sigma_{gated}\left(Lin\_2\left(\sum_{b \in n(a)} C\left(Lin\_1(x_b^{(i)}), r_{ab}\right) + SC\left(x_a^{(i)}\right)\right)\right) \quad (14)$$

where $\sigma_{gated}$ is the equivariant gated nonlinearity, Lin_1 and Lin_2 are equivariant linear layers, $n(a)$ are the neighbors within a specified radius of $a$, $r_{ab}$ is the vector between atom $b$ and the convolution center, and $C$ is the convolution tensor product operation.(14, 43) The self-connection is a linear operation to mix same-irrep channels on a node,

$$SC(x_a) = Wx_a \quad (15)$$

The convolution tensor product operation at center $a$ with neighboring atom $b$ is a tensor product parameterized by weights from a radial basis network, $R$,

$$C(x_b, r_{ab}) = x_b \otimes R(|r_{ab}|)Y(r_{ab}) \quad (16)$$

where $Y$ are the spherical harmonics. Note the maximum order, $l_{max}$, of $Y$ and the cutoff radius is specified by the user. This tensor product follows all the rules of tensor algebra. It permits all the valid tensor operations that yield the desired irrep of the next layer of the network. The radial basis network contains most of the learnable parameters in the convolution (the others come from channel-mixing linear layers before the tensor product). It is a simple multi-layer perceptron (MLP) with one hidden layer acting on a radial basis

$$R(|r_{ab}|) = W_2\left(\sigma_{MLP}\left(W_1(B(|r_{ab}|))\right)\right) \quad (17)$$

where $W_1$ and $W_2$ are weights of the MLP, $B$ are the radial basis functions (we use gaussians), and $\sigma_{MLP}$ is the sigmoid linear unit (SiLU) nonlinearity.

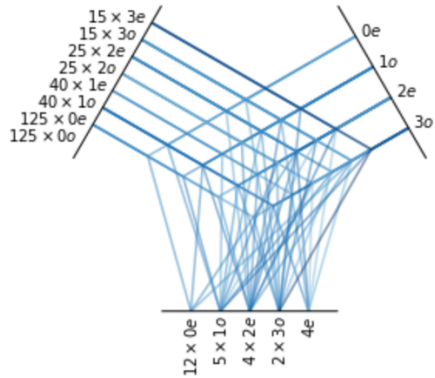

**Figure S1: Illustration of an e3nn Tensor Product.** This shows a fully connected tensor product as implemented in the e3nn convolution of the electron density learning neural network. The example here shows the tensor product from the last hidden layer. The features on atom a, on the left side, combine with the spherical harmonics, on the right side, to give the output features specified by the density-fitting basis set. In the network, each output path is multiplied by its weight from the learned radial function.

The tensor product central to the convolution operation, is illustrated in figure S1. It is responsible for mixing feature and geometric information for each neighboring atom "b" in the environment of atom "a". Each possible mixing "path" in the tensor product is weighted by a scalar provided by the learned radial basis function. At a theoretical level, this fits with the function we are trying to learn; the Hamiltonian in the time-independent Schrödinger equation is a function of the relative strength of various interactions between atoms which are modulated by the atomic geometry (the relative distance vectors between atoms).

The gated nonlinearity is essential to maintaining equivariance in the network. If we were to simply apply a nonlinearity to each feature component, this would break equivariance. (As an example, one cannot multiply each component of a vector separately and expect to maintain the same direction of the vector.) The gated nonlinearity was originally proposed by Weiler and co-workers.(*43*) It works by requiring the convolution operation to output one extra scalar per non-scalar feature (ie. if the hidden irrep has 10 scalars, 5 vectors and 3 rank-2 tensors, the convolution outputs 10+5+3=18 scalars). The extra scalars are fed into a conventional nonlinearity and that output is used to "gate" or scale the magnitude of the non-scalar features. The gated nonlinearity concatenates the separately transformed features in the form,

$$\sigma_{gated}(x) = \sigma_{scalars}(x_{scalars}) \oplus \left( x_{non-scalars} \times \sigma_{gate}(x_{gate-scalars}) \right) \quad (18)$$

where $x$ are the node features, $\sigma_{scalars}$ is the nonlinearity applied to the scalar features, and $\sigma_{gate}$ is the nonlinearity applied to the "extra" scalars. Since only the magnitude of the non-scalars is changed, this nonlinearity preserves equivariance.

**1.2.2. Training Details**

For all the experiments, we set the training set sizes such that each training set would have identical numbers of atoms. This is different than setting the number of structures equal for each training set. Since e3nn learns an atom-centered filter function, this is the appropriate way of comparing across training sets where the number of atoms per structure is different. For the Database of Water Cluster Minima, all clusters with n<7 have fewer than 100 structures, so we choose to start with the n=7 set, which has 469 structures. For this reason, all training sets used in the experiments are small. We note, however, that the learning curves presented in figure 2 of the main text show that one can improve the accuracy systematically by increasing the training set sizes.

# 2 Supplementary Text

## 2.1 Quality of Density Fit

A key component of the electron density learning scheme is the density fitting procedure outlined above. Because the density-fitting basis set we use is not complete, the projection results in a small error relative to the reference density. We set out to analyze the magnitude of this error for the water monomer. Using the equation for density difference, $\epsilon_p$, in the main text, we compute an error of 0.84%. This puts the quality of our density predictions into context. The density fitting approximation is well-tested and widely used in quantum chemistry. The implication is that errors under 1% are acceptable for most quantum chemistry applications. The magnitude of the error in our converged e3nn electron density models is well below this 1% threshold. This means that the model is accurate enough to no longer be the sole, limiting source of absolute error in the electron density. Future work to improve density prediction would demand more complete density-fitting basis set *and* more accurate models trained on more data.

## 2.2 Accuracy of ML electron densities

It is important to put the error in the electron density into context. One relevant comparison is the difference in the electron density between different levels of quantum chemical theory. Bochevarov and Friesner studied the difference in the electron density between various density functionals and full configuration interaction (FCI).(*23*) They used a slightly different metric of density difference,

$$I = \frac{\int d\boldsymbol{r}\big(\rho_{DFT}(\boldsymbol{r}) - \rho_{FCI}(\boldsymbol{r})\big)^2}{\int d\boldsymbol{r}\,\big(\rho_{DFT}(\boldsymbol{r})\big)^2 + \int d\boldsymbol{r}\,\big(\rho_{FCI}(\boldsymbol{r})\big)^2} \qquad (21)$$

The study found that for the small molecules for which FCI calculations were possible (He, $H_2$, LiH, $H_4$), commonly used density functionals like PBE and B3LYP yield density differences on the order of $I=1\times10^{-4}$. This density difference makes a reasonable metric for comparison. DFT methods like PBE and B3LYP are routinely used in quantum chemistry as surrogates when higher-level calculations are not feasible.

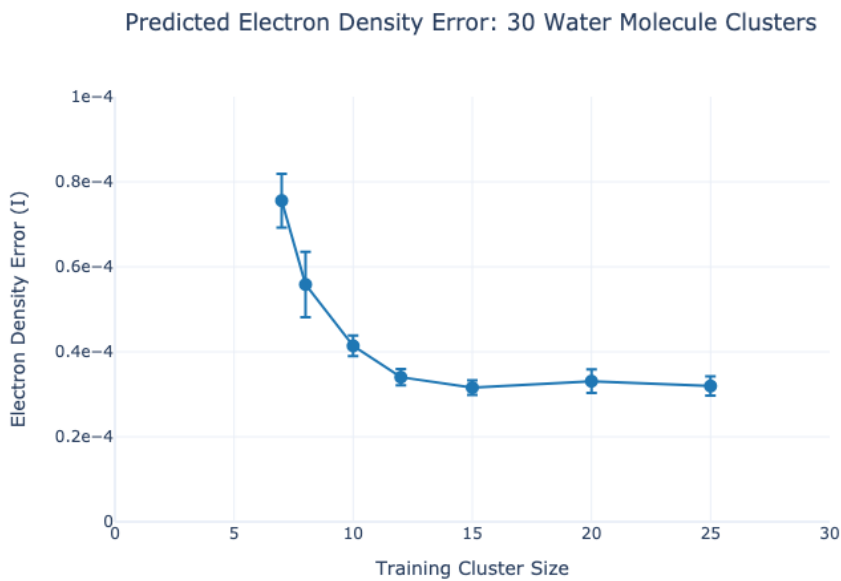

**Figure S2: Density difference measured with *I*.** This shows results from Experiment 1 using *I* from equation 21. Training set cluster size is shown on the x axis and the density difference for each model tested on the n=30 water cluster test set is shown on the y axis.

In figure S2 we plot the density difference error from Experiment 1 in terms of $I$. This is the difference between the ML predicted and PBE0 calculated densities averaged over all structures in the n=30 water molecule test set. This is the exact analog of figure 3 of the main text, but using $I$ instead of $\epsilon_p$. Regardless of training set, the error is always below $1 \times 10^{-4}$. This indicates that the magnitude of error in our e3nn electron density model is lower than the error of DFT methods relative to "exact" FCI. Also note that figure S2 shows the same convergence behavior around a training set cluster size of n=12 as shown in the main text. The $I$ metric for density difference is sensitive to different areas of the density than $\epsilon_p$, indicating that the convergence phenomenon is robust.

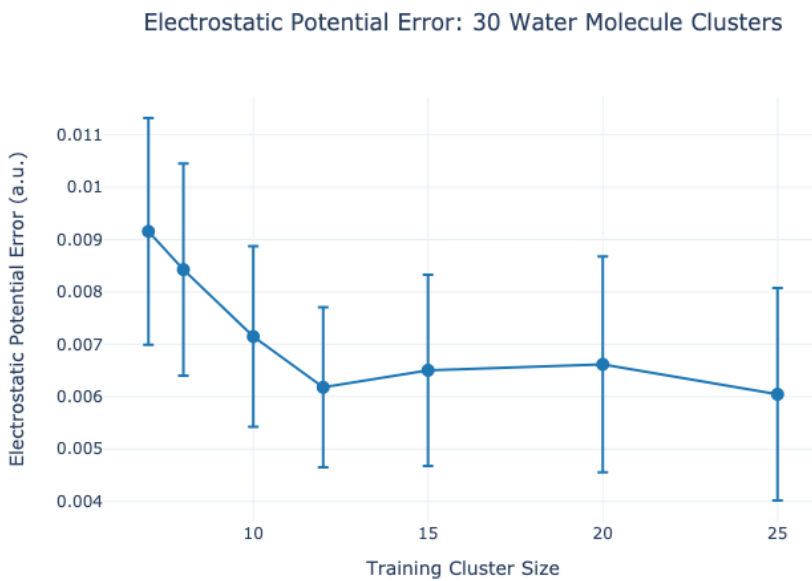

**Figure S3: Electrostatic potential error on n=30 water cluster test set vs. training set cluster size.** The models trained in Experiment 1 are used to calculate the electrostatic potential on the 0.002 e⁻/bohr³ electron density isosurface and compared to the QM (PBE0) result.

Error in the electron density will propagate to error in calculated molecular properties. We set out to determine how the error in our models translates to error in electrostatic potential. We computed the electrostatic potential error at the density isosurfaces of 0.002 e⁻/bohr³ for models trained on training sets n=7 through n=25 on the n=30 test set. The results are plotted in figure S3. The same trend seen for the electron density is seen in the electrostatic potential. The error in the electrostatic potential converges around a training set size of n=12. Moreover, the magnitude of the error in the density isosurface is excellent. Even at the chosen isosurface value of 0.002 e⁻/bohr³, at van der Waals surface of the molecule, the average root mean squared error (RMSE) in the electrostatic potential surface is less than 0.007 a.u.

Previous machine learned electron density models have also achieved high accuracy for electron densities and corollary properties like electrostatic potential. One notable difference of this model is its neural network framework, which allows the model to be systematically improved by training on more data, without increasing the computational cost of inference. Other ML models, notably gaussian process regression, do not have this property. The computational cost of executing the trained model depends on the number of training samples. This underscores the importance of the steep slope of the learning curves illustrated in figure 2 of the main text. The model can be systematically improved by adding more data.

## 2.3 Accuracy of ML forces

A relevant consideration is the accuracy of the ML forces trained in Experiment 2 of the main text. The accuracy of the trained models is excellent. The model trained on the n=25 clusters, achieves a mean absolute error of 0.1 kcal/mol/Å. This error is comparable to state-of-the-art ML force fields, with an error of less than 1% on the 10.2 kcal/mol/Å average total force magnitude of the n=30 test set.(*20*)

Like the ML forces, the long-range Hellmann-Feynman forces calculated from the ML electron densities of Experiment 1 are also very accurate. For example, the long-range Hellmann-Feynman forces calculated from the ML electron density model trained on the n=12 dataset give a mean absolute error of 0.5 kcal/mol/Å on the n=30 test set. This is an error of 3% on the average long-range force magnitude of 16.0 kcal/mol/Å. Note that the average total magnitude of the forces in the long-range case is larger than for the total forces because the structures in the database are close to DFT minima.

## 2.4 Hellmann-Feynman Forces

One of the most appealing features of an ML electron density model is the ability to use it to compute forces directly from the Hellmann-Feynman Theorem. The Hellmann-Feynman Theorem states that the force on any atom can be computed directly from the surrounding electron density.

$$F_i = -\frac{\partial E}{\partial R_i} = \left\langle \Psi \left| \frac{\partial H}{\partial R_i} \right| \Psi \right\rangle + \left\langle \frac{\partial \Psi}{\partial R_i} \left| H \right| \Psi \right\rangle + \left\langle \Psi \left| H \right| \frac{\partial \Psi}{\partial R_i} \right\rangle \tag{22}$$

where $R$ is the nuclear position, $i$ is x, y or z, $H$ is the molecular Hamiltonian, and $\Psi$ is the molecular wavefunction. For variational wavefunctions, the second two terms are zero, which leaves,

$$F_i = \left\langle \Psi \left| \frac{\partial H}{\partial R_i} \right| \Psi \right\rangle = \frac{\partial H}{\partial R_i} \langle \Psi | \Psi \rangle \tag{23}$$

where $\langle \Psi | \Psi \rangle$ is the electron density, $\rho$. Therefore, since the kinetic energy portion of the Hamiltonian does not depend on nuclear positions, the force can be stated as a simple electrostatic integral,

$$F_i = Z_i \left( \sum_{i \neq j} \frac{Z_j(R_j - R_i)}{|R_j - R_i|^3} - \int dr\, \rho(r) \frac{r - R_i}{|r - R_i|^3} \right) \quad (24)$$

where $Z$ are the nuclear charges.

For the Hellmann-Feynman Theorem to be rigorously satisfied, the second and third terms of equation 22 must vanish. In the situation where the basis set explicitly depends on nuclear positions and the basis set is incomplete, however, these terms are not precisely zero. This is the case for the aug-cc-pVTZ basis set used in Experiment 1 of the main text. In this case, there are terms that depend on $\partial \Psi / \partial R$ missing from the Hellmann-Feynman force. These terms are referred to as "Pulay forces". To use Hellmann-Feynman forces for molecular dynamics would require a basis set that is optimized to satisfy the theorem and minimize the Pulay forces.(*44*)

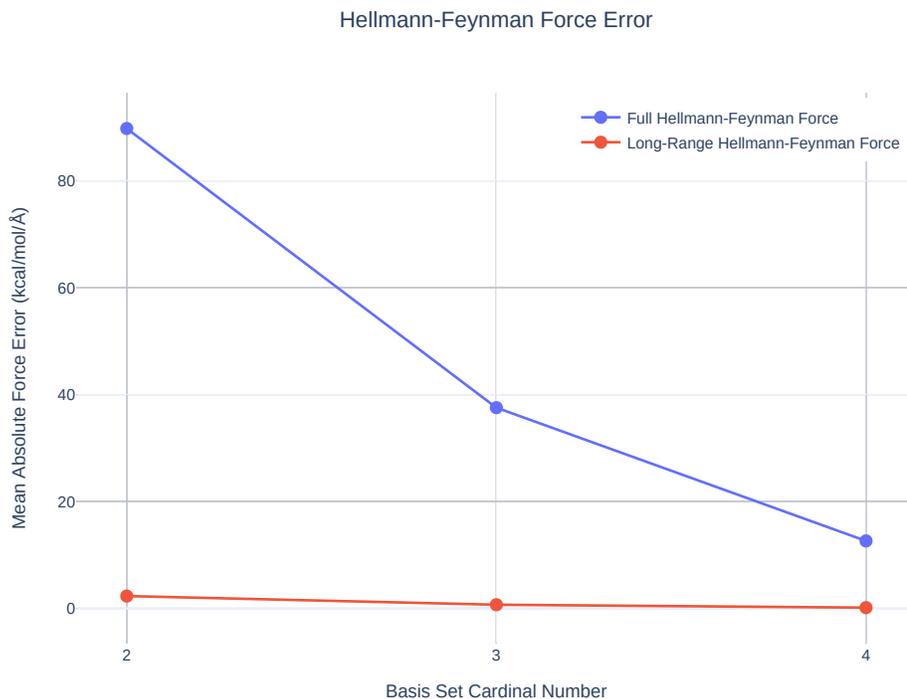

**Figure S4: Convergence of Long-Range Hellmann-Feynman Forces.** For one structure of the n=30 water cluster database, we compute the error in the Hellmann-Feynman forces for various basis sets. The full Hellmann-Feynman force is compared to the Long-Range Hellmann-Feynman force, where the force is computed from contributions of atoms beyond a radius of 7 Å. The basis sets used are aug-cc-pVDZ (2), aug-cc-pVTZ (3) and aug-cc-pVQZ (4). The error is defined relative to a reference aug-cc-pV5Z calculation. The long-range contributions to the Hellmann-Feynman force clearly suffer less basis set error and converge faster with basis set size than the full forces. With the aug-cc-pVTZ basis set used in this study, the contribution of long-range interactions to the error in the Hellmann-Feynman forces about 0.5 kcal/mol/Å.

The purpose of computing Hellmann-Feynman forces in this study is not for molecular dynamics; it is to analyze the contribution to the force from distant atoms. For this task, it is perfectly well-suited. While the total Hellmann-Feynman forces converge slowly with increasing basis set size for conventional Dunning basis sets, such as aug-cc-pVXZ, the contribution to the force from long-range interactions converges quickly. To illustrate this, we computed just the long-range contribution to the Hellmann-Feynman force for all atoms in a member of the n=30 water cluster test set. (We chose to examine just one structure because the aug-cc-pV5Z calculations needed for this test are computationally expensive.) Long-range is defined to be interactions greater than 7 Å. We choose this distance because it roughly coincides with the cluster radius at which convergence occurs for the machine learned electron density. Plotted in figure S4 is the convergence of the Hellmann-Feynman forces with respect to basis set for total and long-range contributions. The long-range contribution to the force clearly converges much faster than the total force. Importantly, it is clearly converged for the aug-cc-pVTZ basis set used for the ML density models trained in Experiment 1. The average error in the long-range contribution to the Hellmann-Feynman force is about 0.5 kcal/mol/Å.

This absolute accuracy makes the long-range Hellmann-Feynman force an excellent tool for examining long-range forces. Experiment 2 shows that an e3nn model trained on atomic forces can continue to improve with information from atoms beyond the long-range radius of about 7 Å. The long-range Hellmann-Feynman forces computed from e3nn models trained on electron densities, however, do not show this same behavior. The accuracy of the long-range forces converges at approximately the cluster size as the density itself. This supports our hypothesis that the electron density is sensitive to a "radius of correlation" that atomic forces are not.

It is worth noting that the magnitudes of the force errors between full forces and long-range forces are not directly comparable. This is largely because the data in this dataset are taken from a database of water cluster minima. Although these are minima on a slightly different potential energy surface than the PBE0/aug-cc-pVTZ surface used here, they are close, which means that the total force on each atom is close to zero. The long-range contributions to the forces, however, are not minima. This means that they will generally be larger in magnitude than the total force, and thus the average error will be bigger. The important thing to note is that the accuracy of the predicted ML electron density long-range Hellmann-Feynman forces are of the same order of magnitude as the expected absolute error in the long-range Hellmann-Feynman force itself. This reinforces the point that the trend with increasing cluster size is the dominant and important feature to take from figure 5 in the main text.

## 2.5 Quantifying convergence with cluster size

To quantify the rate of convergence with cluster size for ML densities and forces, we fit the training cluster size vs. error data with an exponential decay function. We used

$$Error = A + Be^{-C*r} \tag{25}$$

where r is the average maximum distance between atoms in the training cluster, and A, B and C are the fitting parameters. Plotted in figure S5 are the fits for the ML electron density, ML energy+force, and Hellmann-Feynman force models in the main text.

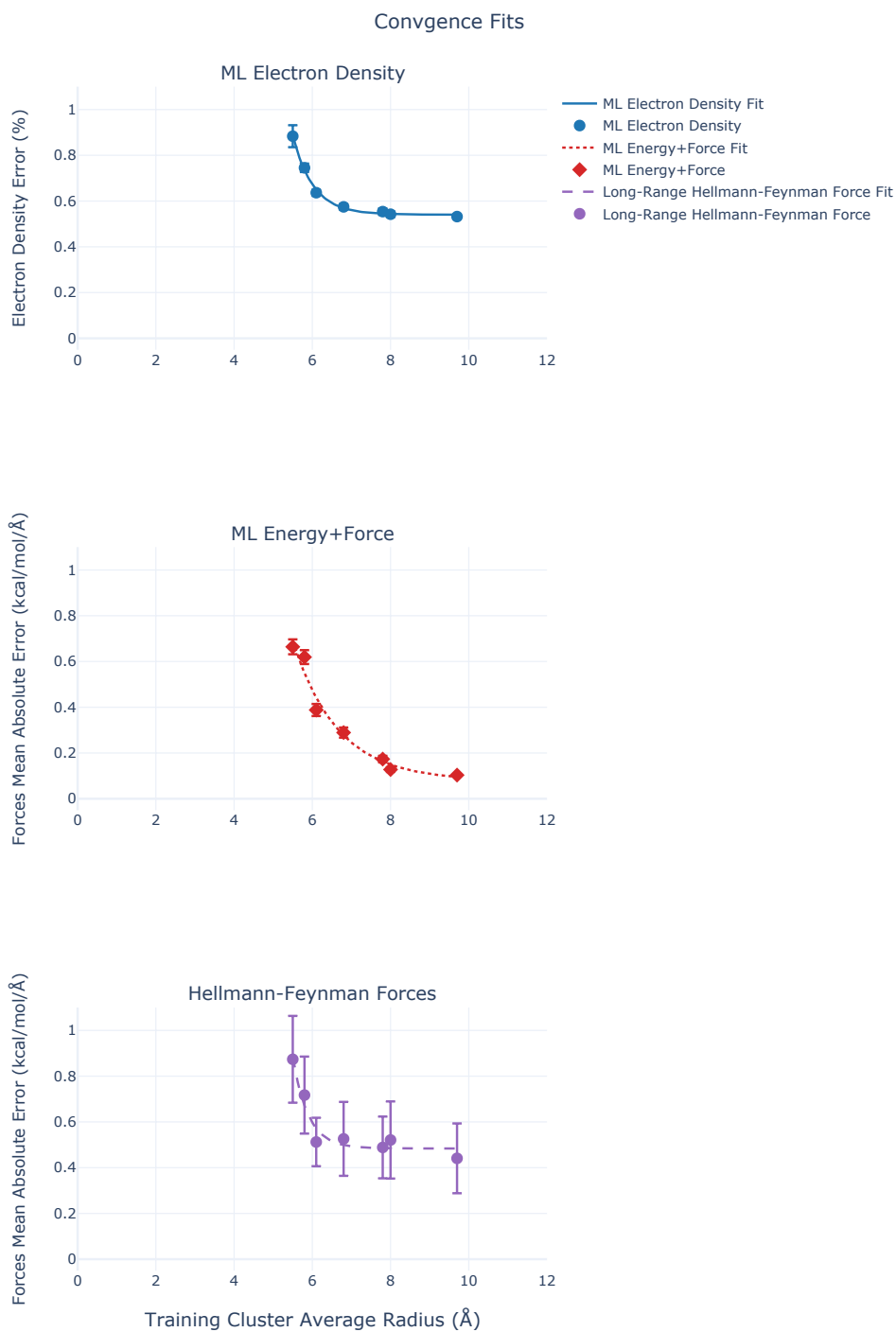

**Figure S5. Training cluster size convergence fits**. Exponential decay functions are fit to the data from Experiments 1 and 2 of the main text. They are plotted as a function of average training cluster radius.

The fits show good agreement with the data in all three cases. The derivatives of these functions are plotted in figure 6 of the main text.